\documentclass[aps,prd,preprint,superscriptaddress,tightenlines,nofootinbib,showpacs]{revtex4}

\usepackage{graphicx}
\usepackage{dcolumn}
\usepackage{bm}

\begin{document}

\preprint{\vbox{\hbox{\hfil CLNS 06/1981}\hbox{\hfil CLEO 06-21}
}}

\title{Measurement of Upper Limits for $\Upsilon\to \gamma + {\cal R}$ Decays}

\author{J.~L.~Rosner}
\affiliation{Enrico Fermi Institute, University of
Chicago, Chicago, Illinois 60637}
\author{N.~E.~Adam}
\author{J.~P.~Alexander}
\author{D.~G.~Cassel}
\author{J.~E.~Duboscq}
\author{R.~Ehrlich}
\author{L.~Fields}
\author{R.~S.~Galik}
\author{L.~Gibbons}
\author{R.~Gray}
\author{S.~W.~Gray}
\author{D.~L.~Hartill}
\author{B.~K.~Heltsley}
\author{D.~Hertz}
\author{C.~D.~Jones}
\author{J.~Kandaswamy}
\author{D.~L.~Kreinick}
\author{V.~E.~Kuznetsov}
\author{H.~Mahlke-Kr\"uger}
\author{P.~U.~E.~Onyisi}
\author{J.~R.~Patterson}
\author{D.~Peterson}
\author{J.~Pivarski}
\author{D.~Riley}
\author{A.~Ryd}
\author{A.~J.~Sadoff}
\author{H.~Schwarthoff}
\author{X.~Shi}
\author{S.~Stroiney}
\author{W.~M.~Sun}
\author{T.~Wilksen}
\author{M.~Weinberger}
\author{}
\affiliation{Cornell University, Ithaca, New York 14853}
% ...
\collaboration{CLEO Collaboration} 
\noaffiliation

\date{April 20, 2007}

\begin{abstract} 

Motivated by concerns regarding possible
two-body contributions to the recently-measured
inclusive $\Upsilon$(nS)$\to\gamma+X$ (n=1, 2, 3)
direct photon spectra,
we report on a new study of exclusive radiative decays of these narrow
$\Upsilon$(nS) resonances into two-body final states ${\cal R}\gamma$, with
${\cal R}$ a narrow resonant hadronic state
decaying into four or more charged particles. 
Such two-body processes are not explicitly addressed in the
extant theoretical frameworks used to calculate the inclusive direct
photon spectra, and must also be explicitly inserted into Monte Carlo
simulations.
Using data collected from the CLEO~III detector at the Cornell Electron
Storage Ring, 
we present upper limits of order $10^{-4}$ for such bottomonium
two-body decays as a function
of the recoil mass $M_{\cal R}$.

\end{abstract}

\maketitle

\section*{Introduction}
CLEO recently extracted $\alpha_s$ from a 
measurement of the direct photon
spectra in $\Upsilon({\rm 1S,2S,3S})\rightarrow gg\gamma$\cite{r:shawn}.  
That extraction was based on a comparison of the $gg\gamma$ width
to the dominant three-gluon width of the narrow bottomonium resonances.
Since the direct photon is observable above background only for
relatively high energies ($E_\gamma\ge E_{\mathrm{beam}}/2$), some model
dependence is inherent in the determination of the 
total $gg\gamma$ rate. Given a 
prescription relating the parton-level rate to 
$\alpha_s$, one can then use that rate to 
determine $\alpha_s$.
To extrapolate beyond the experimentally accessible direct photon energy
region, CLEO relies on 
theoretical parameterizations of the expected photon energy 
spectrum in the $\Upsilon$ system\cite{r:Field,r:SotoGarcia} to
obtain the total direct $\Upsilon\to gg\gamma$ decay width
relative to the dominant $\Upsilon\to ggg$ width. 
The theoretical calculations are generally done at the 
parton level, and therefore
avoid possible resonant
contributions to the photon energy spectrum due to 
two-body decays, e.g., $\Upsilon\to gg\gamma\to\gamma{\cal R}$, with 
${\cal R}$ some resonant hadronic state. Alternatively,
one would like to understand hadronization in 
$\Upsilon\to gg\gamma\to\gamma{\cal R}$, which is measured
experimentally as a two-body process. 
For example, CLEO has recently observed signals in several
low-multiplicity modes\cite{r:pipi05,r:pi0pi006}, and has
presented preliminary
results consistent with the final state $\Upsilon$(1S)$\to\gamma+\eta'$, 
$\eta'\to\pi^+\pi^-\eta$, $\eta\to\pi^+\pi^-\pi^0$\cite{r:potlia}.
We point out, however, 
that the product of (branching ratio)$\times$(efficiency)
for all known exclusive
modes, combined with the observed level of
background, implies a signal yield 
below the statistical sensitivity required to be observed as a distinct
signal in the inclusive spectrum.

From the experimental standpoint, the
presence of possible digluon resonances opposite the photon leads to
``bumps'' in the otherwise smooth predicted theoretical 
photon spectra. 
The inability of the current calculations to directly address
two-body effects, in part, restricts the applicability
of $\Upsilon$ decay models to the region $z_\gamma<0.92$,
with $z_\gamma$ defined as the scaled 
photon energy ($\equiv E_\gamma/E_{\mathrm{beam}}$).  $\Upsilon$
decay models are also not reliable above this point because one of the 
emitted gluons is of such low energy that 
a perturbative calculation cannot be trusted\cite{r:xavier}.  Given that
primary glueball candidates are of order 1 GeV in mass, we expect the
endpoint region of the photon energy spectrum ($z_\gamma>$ 0.92) to be
most susceptible to such contamination. 
Older estimates of $\alpha_s$ based on 
inclusive radiative photon production in $\Upsilon$ decay 
using the BLM\cite{r:Brod-Lep-Mack} prescription
have consistently
yielded values 
smaller than those obtained from different techniques\cite{r:pdg}.
Recently, it has been realized that color octet contributions previously
ignored in the older BLM calculation result in estimates of $\alpha_s$
in excellent agreement with estimates made at the Z-resonance\cite{r:joan07}.
Nevertheless,
subtraction of possibly-enhanced exclusive contributions to the overall
rate would, in principle, result in a lowered estimate for the ratio of rates
$\Gamma(\Upsilon\to gg\gamma)/\Gamma(\Upsilon\to ggg)$ 
and a correspondingly larger
estimate for $\alpha_s$. 
To set the scale,
given that the typical branching fraction for $\Upsilon\to gg\gamma$ is of
order $10^{-2}$, a total resonant
enhancement at the level of $5\times 10^{-4}$ would result in
a $\sim$5\% reduction in the estimate of $\alpha_s$ at the Z-pole.

By comparison,
a large fraction of all $J/\psi\to gg\gamma$ decays have been
identified as two-body\cite{r:pdg}.
Two-body
contamination (and larger
relativistic corrections) also
makes the $J/\psi$ system
somewhat less reliable than the $\Upsilon$ in estimating $\alpha_s$.
This systematic consideration in the $gg\gamma$ analysis motivates
our 
search for radiative decays of the
$\Upsilon$ to resonances: 
$\Upsilon$(nS)$\to \gamma{\cal R}$ (n=1, 2, 3).  We concern ourselves with 
high multiplicity ($\ge 4$ charged tracks) final states,
as we employ
the same hadronic event selection cuts in this analysis that we did
in the $gg\gamma$ analysis \cite{r:shawn}.
We note that
although two-body branching fractions have been observed
for, e.g., $\Upsilon$(1S)$\to\gamma f_2(1270)$ at the level of
$10^{-4}$, the fraction of $f_2(1270)$ decays into $\ge$4 charged
tracks is only $\approx 3\%$\cite{r:pdg}. 

The analysis, in general terms, proceeds as follows. After selecting a
high-quality sample of $e^+e^-$ annihilations into hadrons using the
hadronic event selection cuts of the previous analysis \cite{r:shawn}, 
we construct the
inclusive isolated photon spectra in data taken at both
on-$\Upsilon$-resonance and off-$\Upsilon$-resonance energies (the latter
samples are used for systematic checks of the overall procedure). A two-body
radiative decay of the $\Upsilon$ will produce a monochromatic photon
in the lab frame; the energy of the radiated photon $E_\gamma$
is related to the
mass of the recoil hadron ${\cal R}$ via
$M_{\cal R}=2E_{\mathrm{beam}}\sqrt{1-z_\gamma}$. 
In the case where the intrinsic width of the recoil hadron
is much smaller than the experimental photon energy resolution, the
measured radiative photon energy should be a Gaussian centered at the
energy $E_\gamma$. For a 1 GeV (4.5 GeV) recoil photon, this
implies a recoil resonance with width 
typically narrower than 20 MeV (260 MeV). 
Not knowing \emph{a priori} the mass 
of the hadron
${\cal R}$, we therefore perform a set of fits of the $\Upsilon$(nS) photon
spectrum to a Gaussian signal, centered at a series of $E_\gamma$
values,
and with resolutions corresponding to the known CLEO~III electromagnetic
calorimeter resolution
atop smooth polynomial backgrounds, over the range
$0.2<z_\gamma<1.0$.\footnote{It should be noted that 
``bumps'' in the inclusive photon spectrum can be due not only to
resonant two-body decays but also to 
continuum threshold effects such as 
the crossing of the $c\overline{c}$ threshold.
Given our photon energy resolution,
all processes of the type $e^+e^-\to D^{(*)}{\overline D^{(*)}}$
result in photons relatively close in energy and produce an
apparent enhancement in the region $z_\gamma\sim0.8$.
In the previous analysis\cite{r:shawn}, 
we also identified an excess of photons in
data as $z_\gamma\to1$ ($\equiv E_\gamma/E_{\mathrm{beam}}$). 
Further examination of these events indicated that
they were dominated by continuum
production: $e^+e^-\to\gamma\pi^+\pi^-\pi^+\pi^-$, although
the possibility that the 4-pion state resulted from the decay
of an intermediate resonance ${\cal R}$ was not investigated.}

We construct 95\% confidence level upper limits from these fits 
as a function of recoil mass $M_{\cal R}$,
corrected for the efficiency loss due to the fiducial 
acceptance of the
detector and the event and photon-selection cuts that define our data
sample. In estimating this correction, we assume ${\cal R}$ has spin=0, 
with a  
corresponding $1+\cos^2\theta$ 
angular distribution for the recoil gamma; higher
spins will generally give flatter angular distributions.
Note that the
exact form of the efficiency correction due to the event and photon 
selection cuts varies with the decay final states considered for 
${\cal R}$.  
In the high-momentum region typical of the particles
recoiling against the direct photon, our per-track charged particle
detection efficiency is generally $\ge$90\% over the fiducial
volume.
To be conservative, we derive our 
$z_\gamma$-dependent efficiency correction 
from the decay mode yielding the worst 
reconstruction efficiency.

\begin{figure}[htpb]
\centerline{\includegraphics[width=8cm]{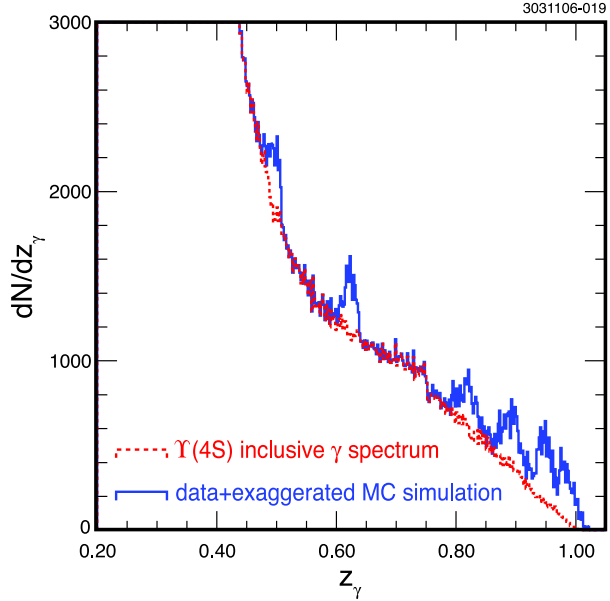}}
\caption{Scaled photon spectrum for
$\Upsilon({\rm 4S})\to \gamma{\cal R}$, ${\cal R}\to 4\pi$ simulations, for
various hypothetical ${\cal R}$ masses.  The lower (dashed) curve is 
$\Upsilon({\rm 4S})$ data, 
while the upper (solid) curve is $\Upsilon({\rm 4S})$ data with signal
Monte Carlo added on top.
The magnitude of 
${\cal B}$($\Upsilon({\rm 4S})\to \gamma{\cal R}$, ${\cal R}\to 4\pi$) 
has been grossly exaggerated for the sake of presentation ($\approx 5 \times 10^{-3}$,
well above any observed radiative decay fraction into $\geq4$ charged tracks).
The six elevations correspond to masses 7.5, 6.5, 4.5, 3.5, 2.5 and,
for the right-most peak, 
the overlap of a ${\cal R}$ of mass $1.5$ GeV and a 
${\cal R}$ of mass $0.5$ GeV, respectively.}
\label{fig:fakeSignal}
\end{figure}

This final efficiency-corrected limit is converted into an
$M_{\cal R}$-dependent branching ratio upper limit 
${\cal B}(\gamma {\cal R})$ by dividing 
the resulting yield by the
calculated total number of resonant $\Upsilon$ events.  For the off-resonance 
running, the distributions are divided by the off-resonance luminosity 
for the sake of comparison between continuum samples.
An example of simulated signal
superimposed on background is given in 
Figure~\ref{fig:fakeSignal} for the 
hypothetical process $\Upsilon({\rm 4S})\to\gamma+{\cal R}$, 
${\cal R}\to\pi^+\pi^-\pi^+\pi^-$.

\section*{Event Selection}
Event selection criteria in this analysis are identical to those imposed
in the previous analysis \cite{r:shawn}. The inclusive photon
spectra are therefore identical to those taken from our previous
analysis as well. The background shape is approximately
exponential in the region $0.2<z_\gamma<1.0$.

\section*{Fitting the Inclusive Photon Spectrum}

To extract the possible magnitude of a two-body
radiative signal, we step along the inclusive photon 
spectrum over the interval $0.2<z_\gamma<1.0$, fitting it to a Gaussian with width
equal to the detector resolution at that value of photon
energy, plus a background
parametrized by a smooth Chebyshyev polynomial.  
We assume that the intrinsic width of the resonance ${\cal R}$ is
considerably smaller than the detector resolution.
Our step size is
determined by the energy resolution of the detector $\sigma_{\rm E}$; we use steps of width 
$\sigma_{\rm E}/2$.

\begin{figure}[htpb]
\centerline{\includegraphics[width=8cm]{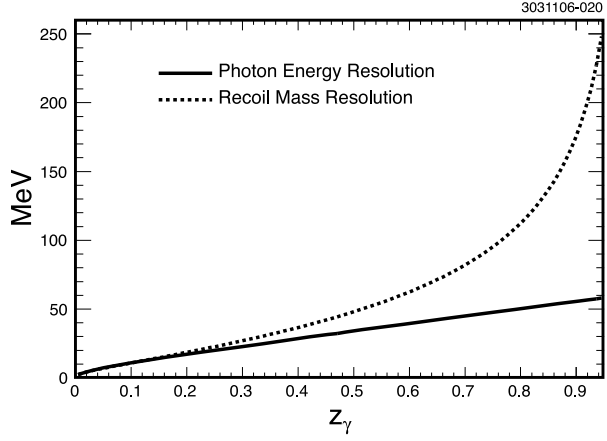}}
\caption{Photon energy resolution (solid)
and recoil mass resolution for 
$\Upsilon$(1S)$\to\gamma{\cal R}$ (dashed).}
\label{fig:xgamma_v_MR}
\end{figure}

For photons in the central barrel
($|\cos\theta_\gamma|<$0.7, with
$\theta_\gamma$ the polar angle of the photon momentum vector relative to the
beam axis) region of the CsI 
electromagnetic calorimeter, the energy resolution over the kinematic interval
relevant to this analysis is of order 2\%.
For two-body radiative decays from the $\Upsilon$(1S), the photon
energy resolution and recoil mass resolution as a function of $z_\gamma$,
are shown in Figure \ref{fig:xgamma_v_MR}. Curves are similar for the
other narrow $\Upsilon$ resonances.

We use a fourth-order Chebyschev polynomial to describe the background.
For each fit, the Gaussian fit area $A(z_\gamma)$ and fit error $\sigma_A(z_\gamma)$
is recorded. 
Note that the fits are highly correlated point-to-point, and
that the bin width is much finer than the detector resolution. 
At each step, we use a
$\pm$10$\sigma_{\rm E}$ fitting window; the background
is expected to be relatively smooth over such a limited interval.  

\section*{Extracting Upper Limits}

\begin{figure}[htpb]
\centerline{\includegraphics[width=8cm]{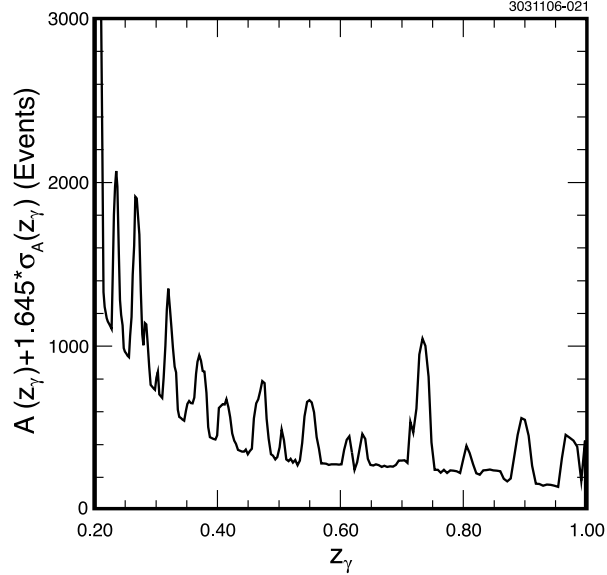}}
\caption{$A(z_\gamma)+1.645\cdot\sigma_A(z_\gamma)$ versus $z_\gamma$ for fits to the 
$\Upsilon$(1S) inclusive photon spectrum where negative points have been mapped to 
$1.645\cdot\sigma_A(z_\gamma)$.  This plot is the upper limit yield, before efficiency correction,
computed from statistical errors only.}
\label{fig:rawUpperLimit1S}
\end{figure}

To convert the $A(z_\gamma)$ distribution obtained from fitting the inclusive photon spectrum
into a 95\% confidence-interval
upper limit, we add $1.645\cdot\sigma_A(z_\gamma)$ point-wise to the 
$A(z_\gamma)$ distribution,
as a function of photon energy.  In this process, since we are interested in
enhancements in the inclusive photon spectrum, all negative areas from the raw fits are
set equal to zero, and the corresponding upper limit
set to $1.645\cdot\sigma_A(z_\gamma)$
at these points.  The resulting contour for the $\Upsilon$(1S) fitting is shown in
Figure~\ref{fig:rawUpperLimit1S}.  The analogous contours for the other $\Upsilon$ spectra look 
similar.

We convert the limits as a function of 
photon energy $z_\gamma$
into a function of a hypothetical resonance recoil mass $M_{\cal R}$.
For the purposes of this conversion, the mean values
for each running period of $E_{\mathrm{beam}}$ 
are used for each data sample; we neglect the MeV-scale variation in
beam energies for a particular run period.  These values are given in Table~I, along with 
the integrated luminosities of the resonance and below-resonance $\Upsilon$ data samples used
in this analysis.

\begin{table}[htpb]
\label{tab:beamEnergies}
\begin{center}
\begin{tabular}{|c|c|c|}\hline
Event Type & $E_{\mathrm{beam}}$ (GeV) & ${\cal L}$ (${\rm pb}^{-1}$) \\
\hline
Resonance $\Upsilon$(1S) & 4.73 & 1076$\pm$11 \\
Resonance $\Upsilon$(2S) & 5.01 & 1189$\pm$12 \\
Resonance $\Upsilon$(3S) & 5.18 & 1228$\pm$12 \\
Resonance $\Upsilon$(4S) & 5.29 & 6456$\pm$65 \\
Below $\Upsilon$(1S) & 4.72 & 188$\pm$2 \\
Below $\Upsilon$(2S) & 5.00 & 396$\pm$4 \\
Below $\Upsilon$(3S) & 5.16 & 158$\pm$2 \\
Below $\Upsilon$(4S) & 5.27 & 2278$\pm$23 \\
\hline
\end{tabular}
\end{center}
\caption{The mean values of average beam energy ($E_{\mathrm{beam}}$) and the integrated luminosity 
(${\cal L}$) for each data sample used in this analysis.}
\end{table}

\begin{figure}[htpb]
\centerline{\includegraphics[width=8cm]{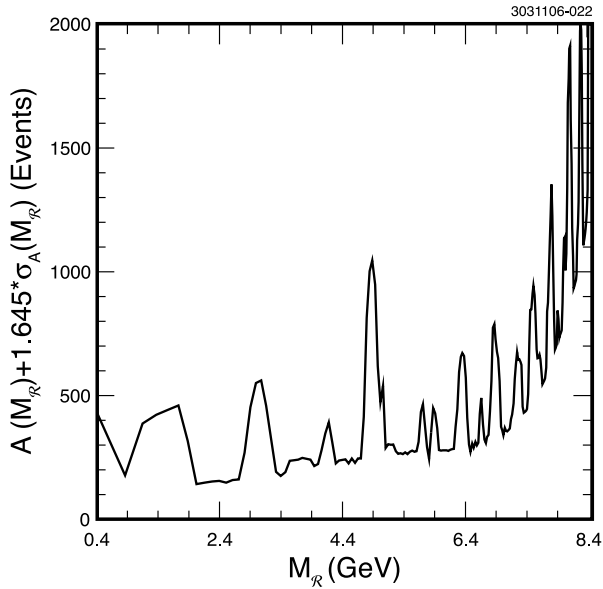}}
\caption{$A(M_{\cal R})+1.645\cdot\sigma_A(M_{\cal R})$ versus $M_{\cal R}$ for fits to the 
$\Upsilon$(1S) inclusive photon spectrum, where negative points have been mapped to 
$1.645\cdot\sigma_A(M_{\cal R})$. Statistical errors included
only.}  
\label{fig:rawRecoilUpperLimit1S}
\end{figure}

The resulting $M_{\cal R}$-dependent contour, 
for the $\Upsilon$(1S), is shown in Figure~\ref{fig:rawRecoilUpperLimit1S}.

\section*{Efficiency Correction}

We consider two efficiency corrections to the upper limit contour:
one due to the fiducial acceptance of the 
detector, and the other due to our event and shower selection cuts.

For photons in the barrel of the detector,
we assume that 
${\cal R}$ is spin zero, 
which corresponds to a $1+\cos^2\theta$ distribution of the 
photons in the two-body decays we are considering; higher
spins will generally give flatter angular distributions \cite{r:pipi05}.  Assuming this 
angular distribution amounts to an $\approx0.6$
uniform angular acceptance efficiency correction factor
to our limit.  

\begin{figure}[htpb]
\centerline{\includegraphics[width=8cm]{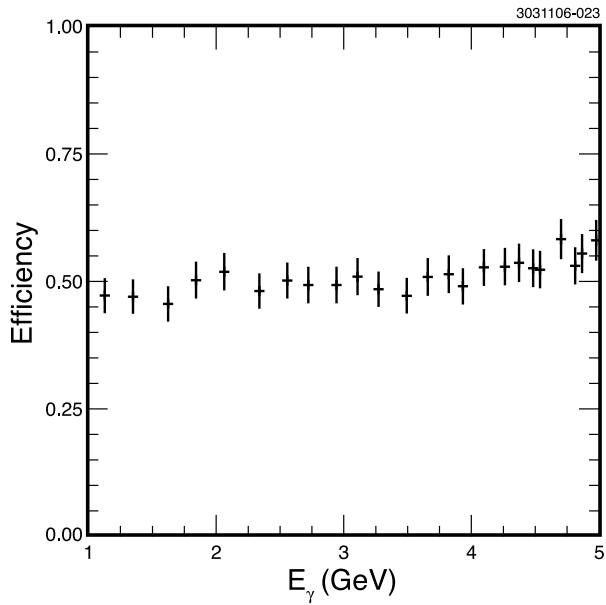}}
\caption{The efficiency for detecting an $\Upsilon\to\gamma+{\cal R}$, 
${\cal R}\to2(K^+K^-)\pi^0$ event as a function of observed photon energy $E_\gamma$ as determined
from a 25000-event Monte Carlo sample.
Each point in this efficiency is obtained from a different 
${\cal R}$ mass hypothesis.  This photon energy-dependent efficiency 
correction distribution is used to point-wise correct our upper limit, where the efficiency between
points in this distribution is estimated by linear interpolation.}
\label{fig:4Kpi0Efficiency}
\end{figure}

In addition to this 
angular acceptance correction,
we assess an efficiency correction due to the CLEO~III detection efficiency.
Not knowing \emph{a priori} what the decay mode of our hypothetical 
resonance ${\cal R}$ will be,
we have
generated 5000-event Monte Carlo samples spanning a wide range of final
state multiplicities (all with $N_{\mathrm{charged}}\ge 4$)
and masses $M_{\cal R}$.
In the interests of producing a conservative upper limit, 
we used this study to choose the mode with the worst average efficiency.
In this manner, we
efficiency-correct, as a function of $M_{\cal R}$ mass, 
our upper limit 
as a function of $z_\gamma$ (or $E_\gamma$) 
before mapping the upper limit into $M_{\cal R}$.
A list of ${\cal R}$ decay modes considered in this analysis and their average 
efficiencies (averaged over the photon energy spectrum from 
$1.0$ GeV $\leq{E_\gamma}\leq4.5$ GeV) is given in Table~II.  
We find that the worst efficiency among the
decay modes considered
was obtained from ${\cal R}\to2(K^+K^-)\pi^0$
(Figure~\ref{fig:4Kpi0Efficiency}).  We therefore use this efficiency 
function to determine our upper limit contours.

\begin{table}[htpb]
\label{tab:eventeff}
\begin{center}
\begin{tabular}{|c|c|}\hline
Event Type & Average Efficiency ($\overline{\epsilon}$)  \\
\hline
${\cal R}\to{K^+K^-\pi^+\pi^-}$ & 0.53 $\pm$ 0.03 \\
${\cal R}\to{K^+K^-\pi^+\pi^-\pi^0}$ & 0.53 $\pm$ 0.02 \\
${\cal R}\to{K^+K^-\pi^+\pi^-\pi^0\pi^0}$ & 0.54 $\pm$ 0.02 \\

${\cal R}\to{K^+K^-p^+p^-}$ & 0.56 $\pm$ 0.02 \\
${\cal R}\to{K^+K^-p^+p^-\pi^0}$ & 0.50 $\pm$ 0.05 \\
${\cal R}\to{K^+K^-p^+p^-\pi^0\pi^0}$ & 0.57 $\pm$ 0.02 \\

${\cal R}\to{p^+p^-\pi^+\pi^-}$ & 0.62 $\pm$ 0.03 \\
${\cal R}\to{p^+p^-\pi^+\pi^-\pi^0}$ & 0.54 $\pm$ 0.05\\
${\cal R}\to{p^+p^-\pi^+\pi^-\pi^0\pi^0}$ & 0.63 $\pm$  0.02 \\

${\cal R}\to{K^+K^-K^+K^-}$ & 0.50 $\pm$ 0.02 \\
${\cal R}\to{K^+K^-K^+K^-\pi^0\pi^0}$ & 0.49 $\pm$  0.02 \\

${\cal R}\to{p^+p^-p^+p^-}$ & 0.67 $\pm$ 0.02 \\
${\cal R}\to{p^+p^-p^+p^-\pi^0}$ & 0.65 $\pm$  0.02 \\
${\cal R}\to{p^+p^-p^+p^-\pi^0\pi^0}$ & 0.63 $\pm$  0.02 \\

${\cal R}\to{\pi^+\pi^-\pi^+\pi^-}$ & 0.59 $\pm$ 0.02 \\
${\cal R}\to{\pi^+\pi^-\pi^+\pi^-\pi^0}$ & 0.65 $\pm$ 0.02 \\
${\cal R}\to{\pi^+\pi^-\pi^+\pi^-\pi^0\pi^0}$ & 0.59 $\pm$ 0.01 \\

${\cal R}\to{\pi^+\pi^-\pi^+\pi^-4\pi^0}$ & 0.57 $\pm$ 0.02 \\
${\cal R}\to{\pi^+\pi^-\pi^+\pi^-6\pi^0}$ & 0.60 $\pm$ 0.02 \\
${\cal R}\to{\pi^+\pi^-\pi^+\pi^-8\pi^0}$ & 0.60 $\pm$ 0.02 \\

${\cal R}\to{K^+K^-K^+K^-K^+K^-}$ & 0.68 $\pm$ 0.04 \\
${\cal R}\to{p^+p^-p^+p^-p^+p^-}$ & 0.53 $\pm$ 0.04 \\
${\cal R}\to{\pi^+\pi^-\pi^+\pi^-\pi^+\pi^-}$ & 0.74 $\pm$ 0.03 \\ 
{\boldmath ${\cal R}\to{K^+K^-K^+K^-\pi^0}$ 
\unboldmath} & {\bf 0.48 $\pm$  0.02} \\ \hline
\end{tabular}
\end{center}
\caption{Average efficiencies for the reconstruction of 
various decay modes 
that could be detected in this analysis, obtained by
fitting the photon energy-dependent reconstruction efficiencies to a straight line
in the interval $1.0$ GeV $<E_\gamma<$ 4.5 GeV (statistical
errors only). Quoted efficiencies correspond to 5000-event Monte Carlo
samples for which photons are generated with
a flat angular distribution within the barrel fiducial
volume of the calorimeter, and therefore
are restricted in geometry. The lowest-efficiency
final state ($K^+K^-K^+K^-\pi^0$ in bold above) is used for setting upper limits.}
\end{table}

\section*{Results}

\begin{figure}[htpb]
\centerline{\includegraphics[width=8cm]{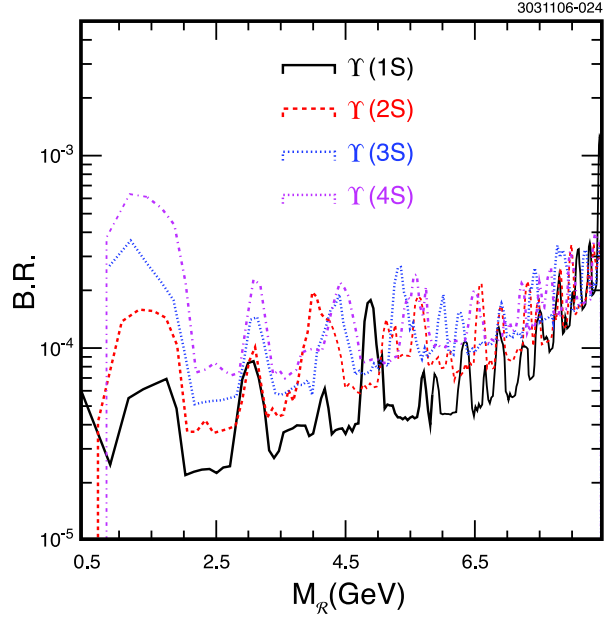}}
\caption{The $M_{\cal R}$-dependent ${\cal B}(\gamma {\cal R})$ upper limit contours obtained for 
$\Upsilon\to\gamma+{\cal R}$, ${\cal R}\to\geq4$ charged tracks for the 
$\Upsilon$(1S), $\Upsilon$(2S), $\Upsilon$(3S)
and $\Upsilon$(4S). Limits are obtained by dividing upper limits on yield by
reconstruction efficiency and number of resonant events, and also
incorporating systematic uncertainties. 
All limits are of order ${\cal B}(\gamma {\cal R})\approx10^{-4}$.}
\label{fig:resonanceLimit}
\end{figure}

To convert the efficiency-corrected upper limit contour into an
upper limit on the two-body radiative branching ratio ${\cal B}(\gamma {\cal R})$,
we simply divide the efficiency-corrected 
upper limit contour by the total
calculated number of $\Upsilon$(nS) decays\cite{r:shawn}, as shown
in Table~III.  For completeness, we also include the results for the $\Upsilon$(4S),
for which the decay width is expected to be nearly saturated by
$\Upsilon$(4S)$\to B{\overline B}$\cite{r:pdg}.  The total number of $\Upsilon$(4S) 
events was obtained by multiplying the total luminosity of our on-resonance $\Upsilon$(4S)-running 
by the well-known $\Upsilon$(4S) on-resonance cross-section (which we calculate from \cite{r:4sxsect}).
The resulting on-resonance 
upper limits ${\cal B}(\gamma {\cal R})$ are shown in Figure~\ref{fig:resonanceLimit}.

\begin{figure}[htpb]
\centerline{\includegraphics[width=8cm]{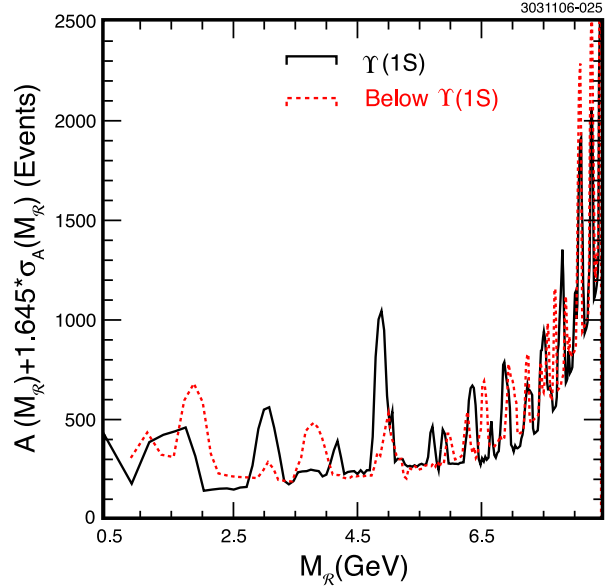}}
\caption{Comparison of the 
$M_{\cal R}$-dependent Gaussian fit area upper limit 
$A(M_{\cal R})+1.645\cdot\sigma_A(M_{\cal R})$  
for the $\Upsilon$(1S) versus the below $\Upsilon$(1S) continuum;
we observe some correlation
between the resonance and the below-resonance structure.  Note that the normalization of the below-resonance upper limit curve is arbitrary and has been adjusted so as to allow an easy visual
comparison.}
\label{fig:limit1sVb1s}
\end{figure}

\begin{figure}[htpb]
\centerline{\includegraphics[width=8cm]{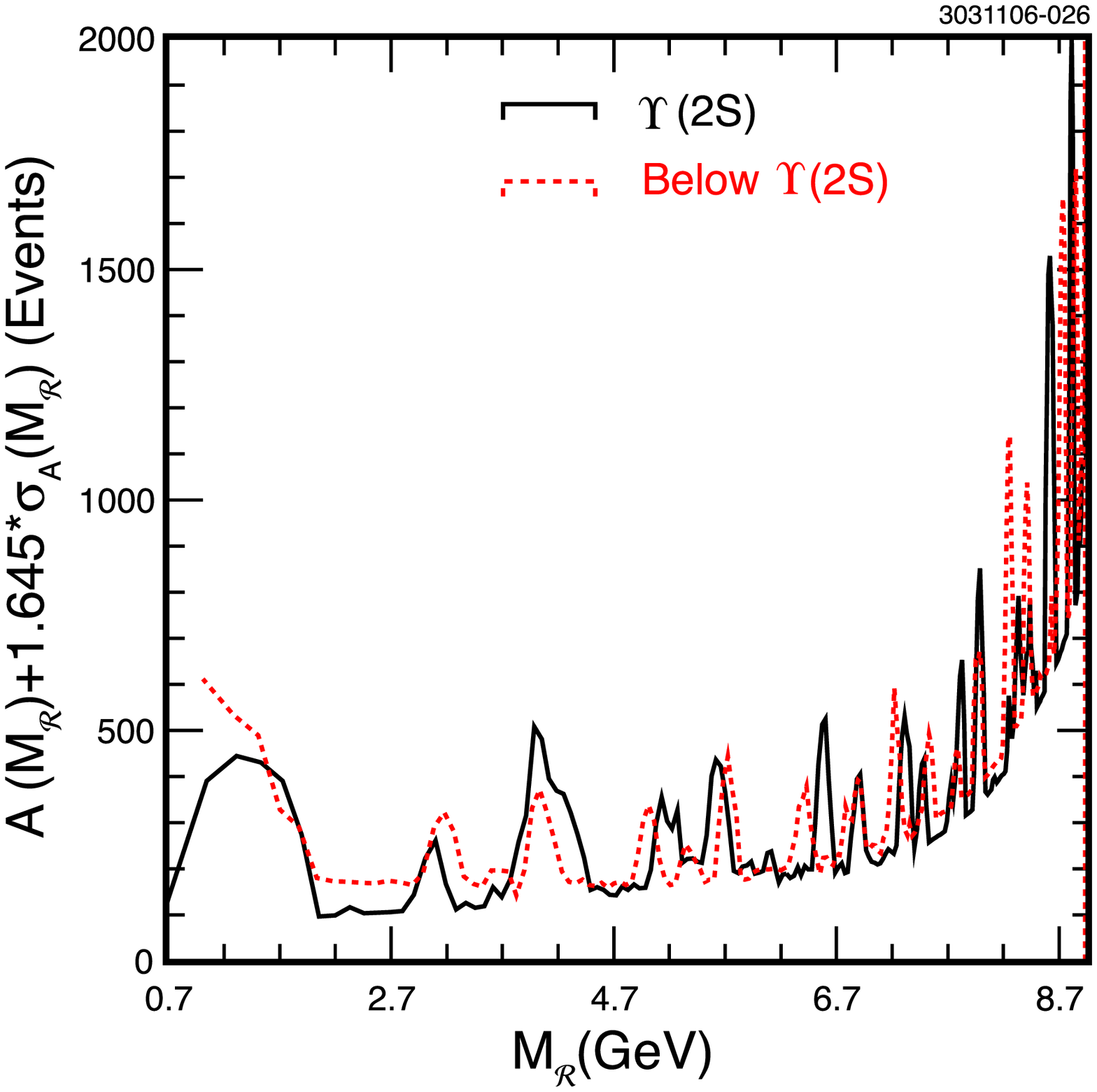}}
\caption{Comparison of the $M_{\cal R}$-dependent Gaussian fit area 
upper limit $A(M_{\cal R})+1.645\cdot\sigma_A(M_{\cal R})$  
for the $\Upsilon$(2S) versus the below $\Upsilon$(2S) continuum;
scaling of the continuum as before.}
\label{fig:limit2sVb2s}
\end{figure}

\begin{figure}[htpb]
\centerline{\includegraphics[width=8cm]{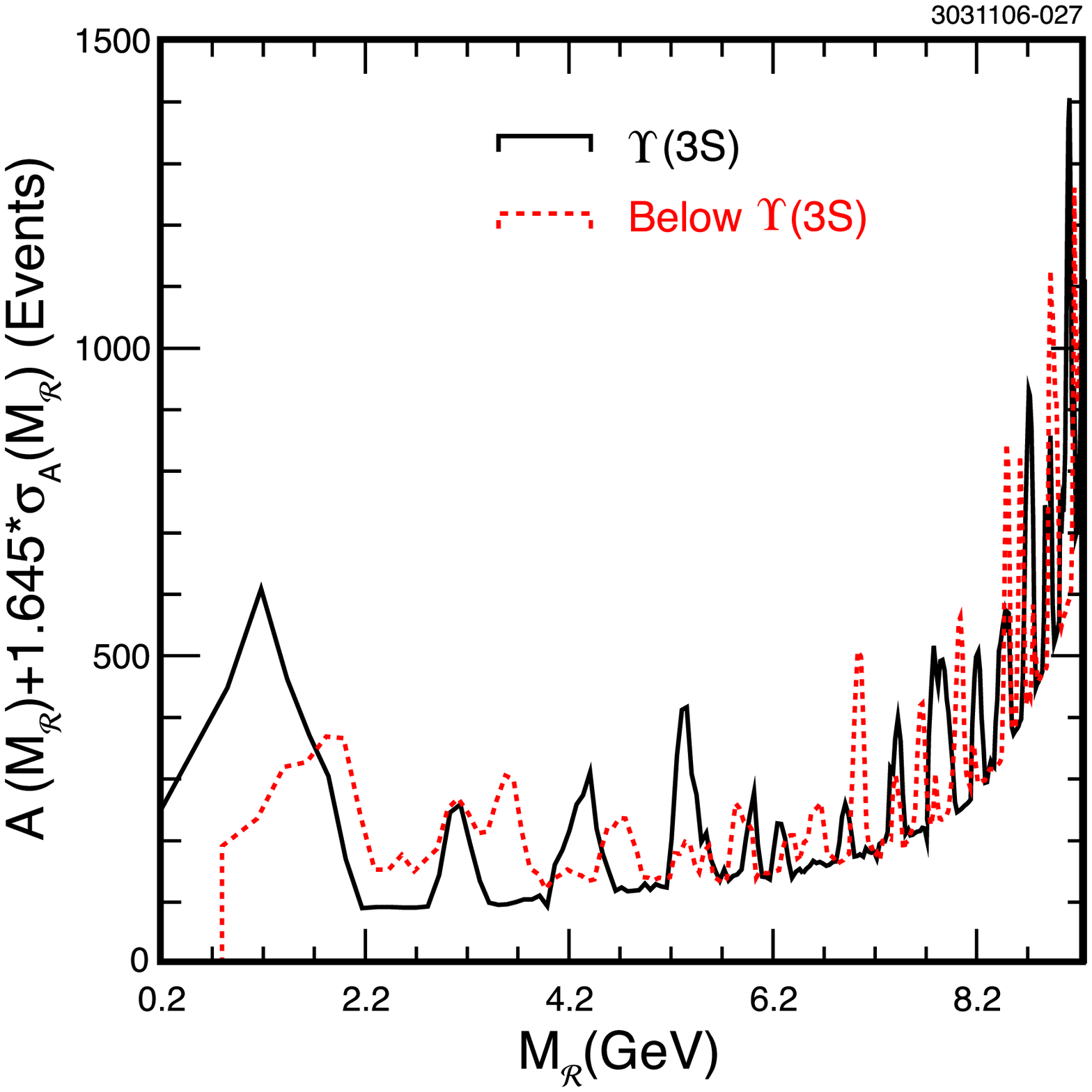}}
\caption{Comparison of the $M_{\cal R}$-dependent Gaussian fit area 
upper limit $A(M_{\cal R})+1.645\cdot\sigma_A(M_{\cal R})$
for the $\Upsilon$(3S) versus the below $\Upsilon$(3S) continuum;
scaling of the continuum as before.}
\label{fig:limit3sVb3s}
\end{figure}

\begin{figure}[htpb]
\centerline{\includegraphics[width=8cm]{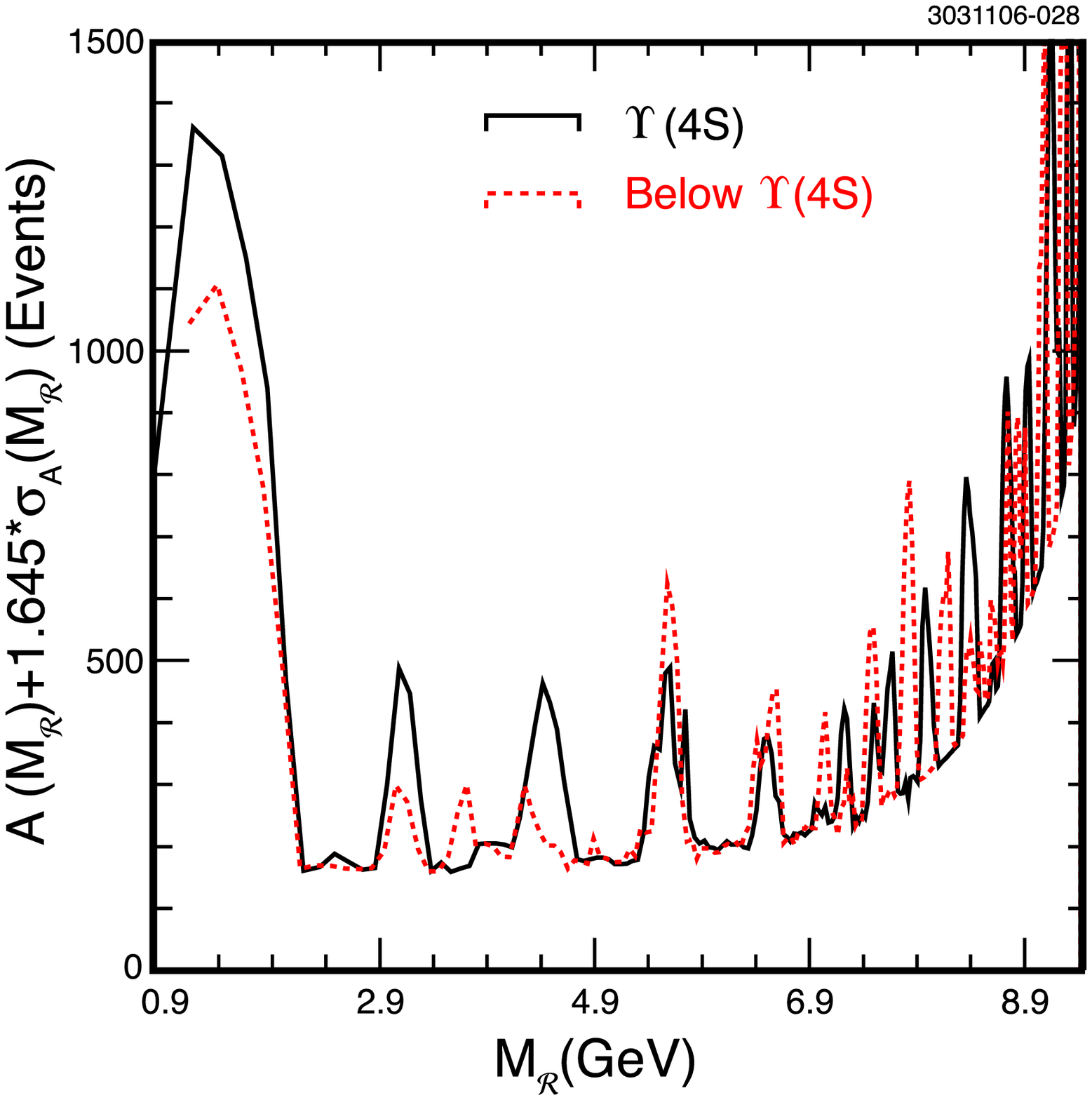}}
\caption{Comparison of the $M_{\cal R}$-dependent Gaussian fit area 
upper limit $A(M_{\cal R})+1.645\cdot\sigma_A(M_{\cal R})$
for the $\Upsilon$(4S) versus the below $\Upsilon$(4S) continuum;
scaling of the continuum as before.}
\label{fig:limit4sVb4s}
\end{figure}

Given the fact that we have not performed a continuum 
subtraction on the on-resonance inclusive photon
spectrum from $\Upsilon$ decays, it is interesting to compare the structure
observed in Figure~\ref{fig:resonanceLimit} with structure observed when we
apply the fitting procedure to continuum data.
Figures~\ref{fig:limit1sVb1s}, \ref{fig:limit2sVb2s}, 
\ref{fig:limit3sVb3s} and \ref{fig:limit4sVb4s}
show the resonances' limits of Figure~\ref{fig:resonanceLimit} 
separately, with the 
corresponding continuum distributions
overlaid for comparison.
We observe 
a partial
correlation between the continuum and resonance spectra, suggesting that
both spectra have large contributions from initial state radiation (ISR)
photons.

\begin{table}[htpb]
\label{tab:eventNumber}
\begin{center}
\begin{tabular}{|c|c|}\hline
$\Upsilon$ Resonance & $N_{\rm total}(\Upsilon$(nS)) $(\times 10^6)$ \\
\hline
$\Upsilon$(1S) & 20.96 $\pm$ 0.06\\
$\Upsilon$(2S) & 8.33 $\pm$ 0.04\\
$\Upsilon$(3S) & 5.24 $\pm$ 0.06\\
$\Upsilon$(4S) & 6.8 $\pm$ 0.2 \\
\hline
\end{tabular}
\end{center}
\caption{The total number of calculated $\Upsilon$(1S), 
$\Upsilon$(2S), $\Upsilon$(3S) and $\Upsilon$(4S) events in our data samples\cite{r:shawn}.}  
\end{table}

\begin{figure}[htpb]\centerline{\includegraphics[width=8cm]{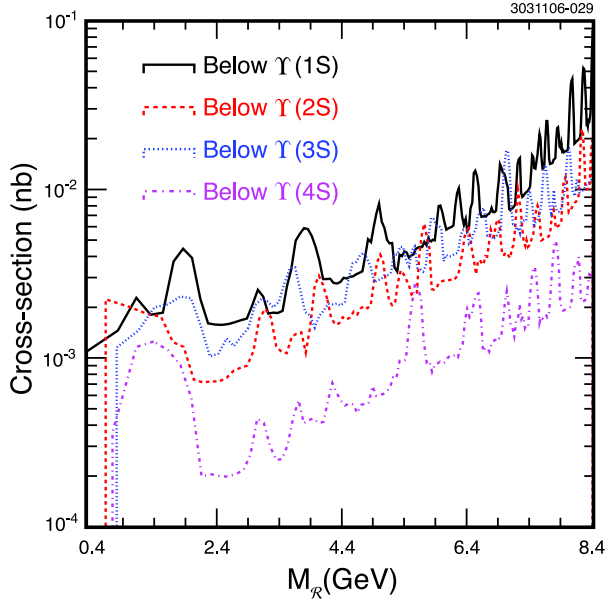}}\caption{$M_{\cal R}$-dependent cross-section upper limit contours obtained for $e^+e^-\to\gamma+{\cal R}$, ${\cal R}\to\geq4$ charged tracks for the below $\Upsilon$(1S), $\Upsilon$(2S), $\Upsilon$(3S) and $\Upsilon$(4S) continua (nb). This plot is obtained by dividing the result of our fitting procedure on the continuum by the off-resonance luminosity. The angular correction here is based on the expected distribution appropriate for continuum initial state radiation. Systematic errors have also been incorporated into these limits.}\label{fig:offResonanceLimit}\end{figure}

Applying our fitting procedure directly to the continuum we can obtain limits on the cross-section for $e^+e^-\to\gamma+{\cal R}$, over the barrel angular acceptance region (Figure~\ref{fig:offResonanceLimit}). It is important to note here that a) the angular distribution for continuum initial state radiation processes is considerably more forward-peaked than the $1+\cos^2\theta$ distribution we have assumed for the resonance; we have therefore applied a correction based on the angular distribution appropriate to ISR, and b) the quantum numbers of particles produced in association with ISR photons are different than those produced in radiative decays of quarkonium resonances. To set the scale of the continuum cross-section sensitivity, the raw ISR cross-section for $e^+e^-\to J/\psi+\gamma$ is  expected to be $\sim$5 pb in the 10 GeV center-of-mass regime. Taking into account the efficiency of our event selection requirements and the strong forward peaking expected for ISR processes, this corresponds to an expected observed cross-section into $\ge$4 charged tracks $\sim10^{-4}$ nb. This value is at the edge of our current statistical sensitivity. 

\section*{Cross-Check}

\begin{figure}[htpb]
\centerline{\includegraphics[width=8cm]{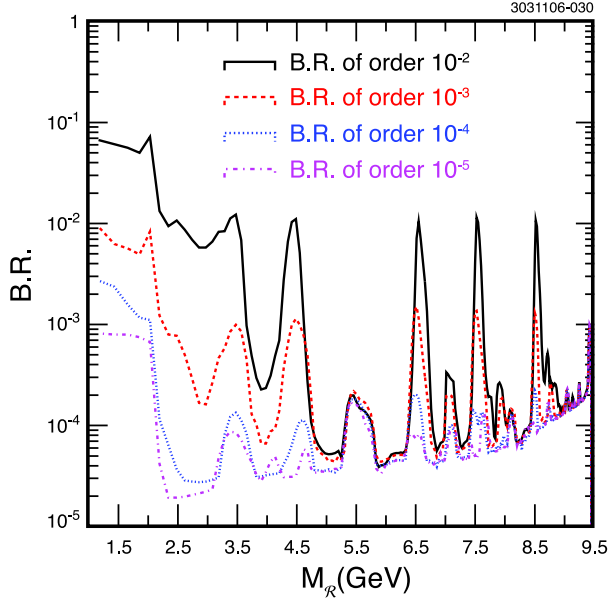}}
\caption{Upper limit contours derived from applying our procedure to 
fabricated Monte Carlo signal spectra.  We reconstruct all input signals withing our sensitivity 
($\approx10^{-4}$) that are within our accessible recoil mass range.}
\label{fig:fake_signal_upper_limit}
\end{figure}

In order to check that we are able to identify a 
signal at a given sensitivity level, we embedded pure Monte Carlo signal 
into data, and 
performed our fitting procedure on the resulting 
distribution to ensure that we recover the correct signal magnitude in 
our branching ratio upper limit.  To do this, 
hypothetical $\Upsilon$(4S)$\to\gamma+{\cal R}$, 
${\cal R}\to\pi^+\pi^-\pi^+\pi^-$ events were embedded into the 
$\Upsilon$(4S) inclusive
photon spectrum with branching ratios of the order of 
$10^{-5}$, $10^{-4}$, $10^{-3}$, 
and $10^{-2}$ under $10$ different $M_{\cal R}$ hypotheses:
$M_{\cal R}=0.6$ GeV, $1.5$ GeV, $2.5$ GeV, $3.5$ GeV, $4.5$ GeV, $6.5$ GeV, 
$7.5$ GeV, $8.5$ GeV, and $9.5$ GeV.
The resulting upper limit contours
derived from applying our procedure to these spectra are show in 
Figure~\ref{fig:fake_signal_upper_limit}.
We reconstruct all signals within our expected 
sensitivity (around $10^{-4}$) that
are within our accessible recoil mass range.

\section*{Systematic errors}
We identify and account for systematic errors as follows:

\begin{enumerate}
\item We account for possible
systematics in our event and shower reconstruction efficiency by using 
the lowest-efficiency final state considered, and by assuming ${\cal R}$ has spin=0.  
The angular distributions for spin=0, 1 and 2 two-body decays have been calculated, and generally
yield flatter distributions for higher spins \cite{r:pipi05}.
\item We have assessed fitting systematic uncertainties by varying the
recoil mass bin width (from 20\% to 50\% of the resolution $\sigma$) and the
order of the background polynomial used to parametrize the background
(from second-order to fifth-order). Observing no statistically significant
variation between these extremes, we have used as defaults
$\sigma$=5 bins and a fourth-order background, based on the goodness
of fit of the pull distributions to a unit Gaussian on the continuum.
\item For continuum
measurements, we assess a uniform 1\% degradation of the limit 
due to the luminosity uncertainty
as calculated in the previous analysis \cite{r:shawn}.
\item For on-resonance measurements,
we degrade the limit uniformly by the uncertainty in the calculated number of total resonant events given in Table~III.  
\end{enumerate}

\subsection*{Summary}
As shown in Figure~\ref{fig:resonanceLimit}, our sensitivity
is of order $10^{-4}$ across the mass range 
corresponding to 0.2$<z_\gamma<$1.0,
well above the tabulated branching ratios for any known
$\Upsilon\to\gamma$+X, X$\to{h^+h^-h^+h^-}$+neutrals
process.
We measure upper limits of:
\begin{center}
${\cal B}(\Upsilon$(1S)$\to\gamma+{\cal R}, {\cal R}\to\geq4$ charged tracks$) < 1.26\times10^{-3}$, \\
${\cal B}(\Upsilon$(2S)$\to\gamma+{\cal R}, {\cal R}\to\geq4$ charged tracks$) < 9.16\times10^{-4}$, \\
${\cal B}(\Upsilon$(3S)$\to\gamma+{\cal R}, {\cal R}\to\geq4$ charged tracks$) < 9.69\times10^{-4}$ \\
\end{center}
for all kinematically allowed 
masses $M_{\cal R}$, under the assumption that ${\cal R}$ is a pseudoscalar.
Constraining $1.5$ GeV $<M_{\cal R}<5.0$ GeV we set a more stringent limit of:
\begin{center}
${\cal B}(\Upsilon$(1S)$\to\gamma+{\cal R}, {\cal R}\to\geq4$ charged tracks$) < 1.78\times10^{-4}$, \\
${\cal B}(\Upsilon$(2S)$\to\gamma+{\cal R}, {\cal R}\to\geq4$ charged tracks$) < 1.95\times10^{-4}$, \\ 
${\cal B}(\Upsilon$(3S)$\to\gamma+{\cal R}, {\cal R}\to\geq4$ charged tracks$) < 2.20\times10^{-4}$.
\end{center}
Additionally, we report these upper limits as a function of the
mass recoiling against the photon, as shown
in Figure~\ref{fig:resonanceLimit}.  

We limit the branching ratio 
for two-body radiative decays to narrow resonances
($<$20 MeV in width)
to be $\le 10^{-4}$.
We conclude that
distortion of the inclusive photon spectrum in our 
previous extraction of $\alpha_s$ due to the possible
contribution of such events 
is negligible.
The possibility of resonances with widths greater than our experimental
resolution has yet to be completely addressed.
Further work on exclusive multiparticle final states (e.g., $\gamma 2\pi^+ 2\pi^-$,
$\gamma 2K^+ 2K^-$, $\gamma K^0K^0$ and $\gamma K^0 K^\pm \pi^\mp$) would help elucidate
the nature of such radiative decays.

\section{Acknowledgments}
We gratefully acknowledge the effort of the CESR staff
in providing us with excellent luminosity and running conditions.
D.~Cronin-Hennessy and A.~Ryd thank the A.P.~Sloan Foundation.
This work was supported by the National Science Foundation,
the U.S. Department of Energy, and
the Natural Sciences and Engineering Research Council of Canada.

\end{document}